\documentclass[12pt]{article}

\usepackage{graphics,epsfig}
\usepackage{amsbsy}
\usepackage{amsfonts}
\usepackage{amsmath}
\usepackage{graphicx}

\def\IR{\mathbb{R}}
\def\IC{\mathbb{C}}
\def\IZ{\mathbb{Z}}
\def\IP{\mathbb{P}}

\def\id{\protect{{1 \kern-.28em {\rm l}}}}

\def\k{\kappa}

\def\fgamma{{\tilde\gamma}}
\def\fpsi{{\tilde\psi}}
\def\fphia{{{\tilde\phi}_1}}
\def\fphib{{{\tilde\phi}_2}}
\def\fphic{{{\tilde\phi}_3}}
\def\tk{{\text{k}}}
\def\p{{\partial}}
\def\nn{\nonumber}

\def \bs {\bigskip}

\def\dalemb#1#2{{\vbox{\hrule height .#2pt
        \hbox{\vrule width.#2pt height#1pt \kern#1pt
                \vrule width.#2pt}
        \hrule height.#2pt}}}

\let\a=\alpha \let\b=\beta \let\g=\gamma \let\d=\delta \let\e=\epsilon
\let\z=\zeta  \let\th=\theta  \let\k=\kappa
\let\l=\lambda \let\m=\mu \let\n=\nu \let\x=\xi \let\p=\pi 
\let\s=\sigma \let\t=\tau   \let\c=\chi 
 \let\vep=\varepsilon
\let\w=\omega      \let\G=\Gamma \let\D=\Delta \let\Th=\Theta \let\L=\Lambda
 \let\P=\Pi \let\S=\Sigma  
\let\C=\Chi \let\W=\Omega
\let\la=\label \let\ci=\cite 
  
\def\nn{\nonumber} \def\bd{\begin{document}} \def\ed{\end{document}}
\def\ds{\documentstyle} \let\fr=\frac \let\bl=\bigl \let\br=\bigr
\let\Br=\Bigr \let\Bl=\Bigl
\let\bm=\bibitem
\let\na=\nabla
\def\tU{{\widetilde U}}
\let\pa=\partial \let\ov=\overline
\def\ie{{\it i.e.\ }}
\newcommand{\be}{\begin{equation}}
\newcommand{\ee}{\end{equation}}
\def\ba{\begin{array}}
\def\ea{\end{array}}
\def\ft#1#2{{\textstyle{{\scriptstyle #1}\over {\scriptstyle #2}}}}
\def\fft#1#2{{#1 \over #2}}
\def\F#1#2{{ F_{#1}^{(#2)} }}
\def\cF#1#2{{ {\cal F}_{#1}^{(#2)} }}

\def\={\, =\, }
\def\+{\, +\, }
\def\-{\, -\, }

\def\R{{\bf R}}
\def\sst#1{{\scriptscriptstyle #1}}
\def\oneone{\rlap 1\mkern4mu{\rm l}}
\def\e7{E_{7(+7)}}
\def\td{\tilde}
\def\wtd{\widetilde}
\def\im{{\rm i}}
\newcommand{\ho}[1]{$\, ^{#1}$}
\newcommand{\hoch}[1]{$\, ^{#1}$}
\newcommand{\bea}{\begin{eqnarray}}
\newcommand{\eea}{\end{eqnarray}}
\newcommand{\ra}{\rightarrow}
\newcommand{\lra}{\longrightarrow}
\newcommand{\Lra}{\Leftrightarrow}
\newcommand{\ap}{\alpha^\prime}
\newcommand{\bp}{\tilde \beta^\prime}
\newcommand{\cB}{{\cal B}}
\newcommand{\cO}{{\cal O}}
\newcommand{\vecx}{\vec{x}}
\newcommand{\vecy}{\vec{y}}
\newcommand{\vecp}{\vec{p}}
\newcommand{\vecq}{\vec{q}}
\newcommand{\tr}{{\rm tr} }
\newcommand{\Tr}{{\rm Tr} }

\newcommand{\cL}{{\cal L}}
\newcommand{\cA}{{\cal A}}
\newcommand{\cD}{{\cal D}}
\def\sst#1{{\scriptscriptstyle #1}}

\def\ve{\varepsilon}
\def\vf{\varphi}
\def\F{\Phi}
\def\wg{\wedge}

\def \foot {\footnote}
\def \bi{\bibitem}

\def \tr {{\rm tr}}
\def \ha {{1 \over 2}}
\def \td {\tilde}
\def \ci{\cite}
\def \N {{\mathcal N}}
\def \ww {\Omega}
\def \const {{\rm const}}
\def \ss {\sum_{i=1}^3 }
\def \t {\tau}
\def\S{{\mathcal S} }
\def \nn {\nu}
\def \XX {{\rm X}}

\def \lra {\leftrightarrow}
\def \vom {{\bar \omega}}
\def \E {{\mathcal  E}} \def \J {{\mathcal  J}}
\def \YY {{\rm Y}}

\def \d {\del}
\def \rJ {{J}}
\def \sms {sigma models\ }
\def \sm {sigma model\ }
\def \L {\Lambda}
\def \gl {\ell}
\def \tr {{\rm tr\ }}
\def\z{\zeta}
\def\zi{\zeta_1}
\def\zii{\zeta_2}
\def\K{\mbox{K}}
\def\eE{\mbox{E}}   \def \vt {\vartheta}
\def \vr {\varrho}
\def \wup {w}

\def\dg{\dagger}
\def\a{\alpha}
\def\b{\beta}
\def\e{\varepsilon}
\def\p{\phi}
\def\ap{\alpha^\prime}
\def\I{{\cal I}}

\def\R{{\bf R}}
\def\Z{{\bf Z}}
\def\C{{\bf C}}
\def\P{{\bf P}}
\def\xb{{\bar X}}
\def\Tr{{\rm  Tr}}
\def\tr{{\rm  tr}}

\def \del{\partial}
\def \a {\alpha}
\def \aa {{\a'}}
\def\g{\gamma}
\def\s{\sigma}
\def\z{\zeta}
\def\zi{\zeta_1}
\def\zii{\zeta_2}
\def\ov{\over}

\def\I{{\cal I}}
\def\J{{\mathcal J}}
\def \ok {{1\ov \k}}
\def\LL{{\mathcal L }}
\def \jL {{J}}
\def \om {\omega}
\def \cL {{\mathcal L}} \def \cH {{\mathcal H}}
\def\E{{\mathcal E}}
\def\w{\omega}
\def\b{\beta}
\def\l{\lambda}
\def\eps{\epsilon}
\def\vep{\varepsilon}
\def \De {{\mathcal D}}

 \def \cV {{\cal V}}
\def  \Jt {  {J}_{\rm tot}    }

\def \k {\kappa}
\def\foot{\footnote}
\def \four{{\textstyle {1\ov 4}}}
 \def \third { \textstyle {1\ov 3
}}
\def\det{\hbox{det}}
\def \ci {\cite}

\def \foot {\footnote}
\def \bi{\bibitem}

\def \tr {{\rm tr}}
\def \ha {{1 \over 2}}
\def \tid {\tilde}
\def \vv {{\rm v}}
\def \tl {{\tilde \l}}
\def \XX {{\rm X}}
\def \ta {{\tilde \a}}
\def \fo { {1\ov 4}}
\def \ep {\epsilon}
\def \inti {{\int^{2\pi}_0 {d \sigma \ov 2 \pi}}}

\def \d {\partial}
\def \K {{\rm S}}
\def \el {\ell}
\def \Tr {{\rm Tr}}
\def \P {\Phi}
\def \l  {\lambda}
\def \tl {{\tilde \l}}
\def \bl {{\tilde \l}}
\def \const {{\rm const}}
\def \V {v}

\def \bv {v^*}
\def \vv {{\rm v}}
\def \LL {{\mathcal L}}
\newcommand{\PV}[1]{P_{\!\!_{V_{#1}}}}

\def \bL {\ell}
\def \M {{\mathcal M}}
\def \N {{\mathcal N}}
\def \S {{\rm S}}
\def \vn {\vec n}
\def \tl {\td \l}
\def \td {\tilde}
\def \Prod {\Pi}
\def \O {{\mathcal O}}
\def \Q {{\rm  Q}}
\def \D {\Delta}
\def \N {{\mathcal N}}
\def\tN{{\tilde N}}

\def \m {\mu}
\def \vs {\vec \s}
\def \ie {i.e.}

\def \cD {{\cal D}}

\def  \le  {\l_{\rm eff}}

\def \rS {{\rm S}}
\def\as{{\a}}
\newcommand{\bra}[1]{\mbox{$\langle #1 |$}}
\newcommand{\ket}[1]{\mbox{$| #1 \rangle$}}

\newcommand{\auth}{AUTHORS}

\def\thb{\bar{\theta}}
\def\Thb{\bar{\Theta}}
\def\barp{\bar{p}}
\def\barq{\bar{q}}
\def\barc{\bar{c}}
\def\bard{\bar{d}}
\def\e{\epsilon}

\def \bi{\bibitem}
\def \la {\label}

\def \l {\lambda}
\def\foot{\footnote}
\def \tl  {{\tilde \l}}
\def \sql {{\sqrt \l}}
\def \adss {$AdS_5 \times S^5~$ }
\newcommand{\rf}[1]{(\ref{#1})}
\def \ov {\over}

\def\th{\theta}
\def\Th{\Theta}
\def\vth{\vartheta}

\def\vth{\vartheta}

\def\ra{\rightarrow}
\def\N{{\cal N}}
\def\F{{\cal F}}
\def\cc{\circ}
\def\eqv{\equiv}

\def\ni{\noindent}
\def \ha{{1\ov 2}}
\def \bw {{\rm w}}

\def\r{{\rm r}}

\def \cT {{\cal T}}
\def \no {\nonumber}

\def \J {\mathcal{J}}
\def \del {\partial}

\def \bps {{\bar \psi}}
\def \sqbl {\sqrt{\bar \lambda}}

\def\dF{\dot{F}}
\def\dG{\dot{G}}
\def\df{\dot{f}}
\def \E {{\cal E}}

\def \S {{\cal S}}
\def \J {{\cal J}}

\def\ms{\mathcal{S}}
\def\mj{\mathcal{J}}
\def\soj{\fr{\ms}{\mj}}
\def \R {{\bf R}}
\def \om {\omega}
\def \tH {\widetilde H}
\def \bE {\bar E}
\def \x {{\cal X}}

\def \hV {{\hat V}}

 \def \bb {\bar \beta}
\def \W {{\cal E}}

\def \bi{\bibitem}
\def \la {\label}

\def \l {\lambda}
\def\foot{\footnote}
\def \tl  {{\tilde \l}}
\def \sql {{\sqrt \l}}
\def \sqtl {{\sqrt {\tilde \l}}}

\def \HH {{\rm E}}
\def \cS {{\cal S}}
\def \cL {{\cal L}}

\def \adss {$AdS_5 \times S^5$\ }

\def \D {\Delta}
\def \thet {\theta}
 \def \t {\tau}
 \def \p {\phi}
 \def \r {\rho}
 \def \rN {{\rm N}}
 \def\tw{{\tilde w}}
 \def\hJ{{J}}
 \def\hw{{w}}
 \def\hl{{\lambda}}
 \def\hth{{\theta}}
 \def\NN{{\cal N}}
 \def \bv {{ \bar w}}
\def \vn {{\vec n}}
\newcommand{\sfrac}[2]{{\textstyle\frac{#1}{#2}}}
\def \bl {{ \bar \lambda}}
\def \bp {{\bar p}}
\def \bu {{\bar u}}
\def \sha {\sfrac{1}{2}}
\def \w {\omega}
\def \ov {\over}
\def \vl { \vec \ell}
\def \varpi {{\rm w}}
\def \OO {{\cal O}}
\def \bG {\bar \G}

\def \c {\gamma}

\def \ss {{\rm s}}

\def \ve {\varepsilon}
\def \pa{\partial}
\def \I {{\cal I}}
\def \LL {{\cal L}}
\def \ep {\epsilon}
\def \R {{\rm R}}
\def \tilt {{\tilde t}}

\def\pic #1#2{\hbox{\lower#1pt\hbox{~\mbox{\epsfxsize=20truemm \epsffile{#2}}}}}

\def\pic #1#2#3{\hbox{\lower#1pt\hbox{~\mbox{\includegraphics[scale=#3]{#2}}}}}

\def \bss {\bigskip}

\def \KK {{\rm K}}

\def\tx{{\tilde x}}
\def\ty{{\tilde y}}
\def\tf{{\tilde f}}
\def\tz{{\tilde z}}
\def\tpo{{\tilde \phi}_1}
\def\tpt{{\tilde \phi}_2}
\def\hx{{\hat x}}
\def\hz{{\hat z}}

\def\IR{\mathbb{R}}
\def\IC{\mathbb{C}}
\def\IZ{\mathbb{Z}}
\def\IP{\mathbb{P}}

\def\be{\begin{eqnarray}}
\def\ee{\end{eqnarray}}

\def\id{\protect{{1 \kern-.28em {\rm l}}}}

\def\bfsigma{{\boldsymbol{\sigma}}}

\def\lin{{-\!\!\!-\!\!\!\!-\!\!\!\!-\!\!\!-\!\!\!-}}

\def\pint{{-\!\!\!\!\!\!\int}}

\def \bt {\bar\theta}
\def \te {\theta}

\def \cc {{\rm f}}
\def \d {\delta}

\def \cL {{\cal L}}
\def \S  {{\rm S}}
\def \pp {{q}}
\def \vt {\vartheta}
\def \mm {{\cal  \ell}}
\def \Z {{\cal Z}}
\def \pa {\partial}
\def \C {{\cal C}}
\def \be {\bea}
\def \ee {\eea}
\def \c {\gamma}  \def \d {\delta}

\def \DD {{\rm D}}
\def \chii {\varepsilon}
\def \th {\theta}

 \def \t {\tau}
 
\def \beq {\be}
\def \eeq {\ee}
\def \beqa {\bea}
\def \eeqa {\eea}

\def\tx{{\tilde x}}
\def\tz{{\tilde z}}
\def\hx{{\hat x}}
\def\hz{{\hat z}}

\def\fx{~{\tilde x}}
\def\fy{~{\tilde y}}
\def\fpo{~{{\tilde \varphi}_1}}
\def\fpt{~{{\tilde \varphi}_2}}
\def\ff{~{{\tilde \phi}}}

\def\Tr{{\rm Tr}}
\def \cA {{\cal A}}

\def\IR{\mathbb{R}}
\def\IC{\mathbb{C}}
\def\IZ{\mathbb{Z}}
\def\IP{\mathbb{P}}

\def\id{\protect{{1 \kern-.28em {\rm l}}}}

\def\be{\begin{eqnarray}}
\def\ee{\end{eqnarray}}

\def\MMk{\kappa_f}
\def\fgamma{{\tilde\gamma}}
\def\fpsi{{\tilde\psi}}
\def\fphia{{{\tilde\phi}_1}}
\def\fphib{{{\tilde\phi}_2}}
\def\fphic{{{\tilde\phi}_3}}
\def\tk{{\text{k}}}
\def\p{{\partial}}
\def\nn{\nonumber}

\makeatletter
\renewcommand\section{\@startsection {section}{1}{\z@}%
                                   {-3.5ex \@plus -1ex \@minus -.2ex}%
                                   {2.3ex \@plus.2ex}%
                                   {\normalfont\large\bfseries}}
\renewcommand\subsection{\@startsection{subsection}{2}{\z@}%
                                   {-3.25ex\@plus -1ex \@minus -.2ex}%
                                   {1.5ex \@plus .2ex}%
                                   {\normalfont\normalsize\bfseries}}
\makeatother

\def\b{{\rm b}} 
\def\d{{\rm d}} 
\def\aa{{\rm a}}
\def\uo{{\sigma_0}}
\def\ut{{\sigma_1}}
\def\uu{{\sigma}}

\def\tx{{\tilde x}}
\def\ty{{\tilde y}}
\def\txx{{\xi}}
\def\tyy{{\eta}}
\def\tfpo{{{\varphi}_1}}
\def\tfpt{{{\varphi}_2}}
\def\tphi{{\phi}}

\def \KK {{\rm K}}
\def \b {\beta}
\def \bs {\bigskip}
\def \c {{\rm a}}
\def \del {\partial} 
\def \d {\partial}
\def \s {\sigma}

\def\eps{{\epsilon}}

\def\Li{{\rm Li}}

 \def \L {{\cal L}}
 \def \sq {\sqrt 2}

\begin{document}
\overfullrule=0pt
\parskip=2pt
\parindent=12pt
\headheight=0in \headsep=0in \topmargin=0in \oddsidemargin=0in

\vspace{ -3cm}
\vspace{-1cm}

\rightline{Imperial-TP-AT-2007-3}

\begin{center}
\vspace{0.1cm}
{\Large\bf
Strong-coupling expansion  of cusp anomaly\\
\vspace{0.2cm}
from quantum superstring  
\vspace{0.3cm}
   }

 \vspace{.2cm} {
  R. Roiban$^{a,}$\footnote{radu@phys.psu.edu}
 and A.A.
 Tseytlin$^{b,}$\footnote{Also at
 Lebedev  Institute, Moscow.
  tseytlin@imperial.ac.uk
 }}\\
 \vskip 0.03cm

{\em 
$^{a}$Department of Physics, The Pennsylvania  State University,\\
University Park, PA 16802 , USA\\
\vskip 0.08cm $^{b}$  Blackett Laboratory, Imperial College,
London SW7 2AZ, U.K. }

\end{center}

 \begin{abstract}
We consider  the world surface  in $AdS_5$ that ends 
on two intersecting  null lines  at the boundary. 
The corresponding  superstring partition function 
describes the expectation  value of the Wilson line with a  null cusp
in dual large $N$ maximally 
supersymmetric gauge theory and thus  determines 
the cusp anomaly  function $f(\l)$ of the gauge  coupling $\l$ 
or the string tension ${\sql \ov 2 \pi}$. 
The first two  coefficients in its  strong-coupling 
or string inverse tension expansion 
were determined  in hep-th/0210115 ($\c_0={1}$)
and in 	arXiv:0707.4254  ($\c_1=-{3 \ln 2  }$).
Here we find  that   the 2-loop  coefficient is    $\c_2 = -  \KK$   where $\KK$ 
is the Catalan's constant. This is in agreement (expected on the 
 general grounds) 
with the previous results 
 for $f(\l)$ as the   coefficient of $\ln S  $ term 
in the energy of the closed  spinning  string in $AdS_5$. 
The string theory value for   $\c_2$ is in
agreement with the numerical result
 in hep-th/0611135 and the  recent analytic result in
  arXiv:0708.3933 for the coefficients in strong-coupling 
solution of the BES equation.
We explicitly verify   the cancellation of all 2-loop
logarithmic divergences
thus   demonstrating
the quantum consistency of the  \adss superstring action at this order.
We also  discuss 
the structure of the three  and higher string 
loop corrections to the cusp anomaly function 
giving a 2d QFT  diagrammatic 
interpretation to the  result  of  arXiv:0708.3933
for the  solution  of the BES equation 
following from the  Bethe ansatz prescription for the spectrum
of the theory.
\end{abstract}
\newpage

\renewcommand{\theequation}{1.\arabic{equation}}
 \setcounter{equation}{0}

\setcounter{equation}{0} \setcounter{footnote}{0}
\setcounter{section}{0}

\section{Introduction}

Anomalous  dimension of minimal twist  large spin single trace operator
or anomalous  dimension of a Wilson line 
with a null cusp \ci{cusp}
 was  a subject of  much attention 
in the context of the  AdS/CFT duality for several years starting
 with the seminal work of 
\ci{gkp} (see also \ci{ft1,kru,mak}). In the planar limit  this dimension  is 
a function  $f(\l)$  of the `t Hooft coupling $\l$  or of the \adss   string
 tension $ { \sql \ov 2 \pi}$. 
 Finding this function exactly would be an important progress. 
A series of recent developments based on the apparent 
 integrability of the theory
 culminated in a suggestion \ci{bes} of an  integral  equation 
 that,  in principle,   determines  $f(\l)$  for
 any value of $\l$. 

To check the consistency of this equation and thus of the underlying 
asymptotic Bethe ansatz  it is important compare its prediction with that
of the quantum superstring theory in \mbox{\adss.}  The perturbative string theory or the 
strong-coupling  expansion of $f(\l)$ can be written as 
\be \la{f}
f(\l) = { \sql \ov \pi}\ \big[\c_0 + { \c_1 \ov \sql} + { \c_2 \ov (\sql)^2 } +
 { \c_3 \ov (\sql)^3 } 
+ ... \big] \ , \ee
where the tree-level \ci{gkp} and the 1-loop \ci{ft1} 
superstring predictions 
are
\be \la{ff}
 \c_0=1  \ , \ \ \ \ \  \ \ \ \ \ \ \c_1 = - 3 \ln 2    \ . \ee 
The computation of the 2-loop superstring coefficient  was initiated in 
\ci{rtt}\foot{Note that in 
 the notation of \ci{rtt}  $a_k = { 1 \ov \pi} \c_k$.}  
where it was found to be expressed in terms  of  the Catalan's constant 
$\KK= \sum^\infty_{n=0} { (-1)^n \ov (2n+1)^2} \approx 0.9159.$

The expansion of the BES \ci{bes} 
equation at strong coupling turned out to be a non-trivial problem
\ci{ben,lip,benn,kos,bec}.\foot{See  also \ci{bel} for a potentially
 important alternative approach based on Baxter equation. 
 A strong-coupling 
 interpolation of the sum of 
 few leading 
 perturbative gauge-theory coefficients which  appears  to be  in good agreement 
 with the string results  \rf{ff}   was discussed in \ci{lipa,dix,lip}.}
 The results for the three  leading $\c_n$ coefficients 
 \rf{ff}  were first found
only  numerically  \ci{ben} ($\c_0$ was later computed exactly 
\ci{benn}).\foot{$\c_1$ 
was also computed \ci{char} 
 from  the ``string'' version of the 
  Bethe ansatz, i.e.  with the magnon scattering 
  phase taken in the  strong-coupling expanded form 
  \ci{bhl}.}
The numerical result for the third coefficient  found in \ci{ben} was 
$\c_3 \approx - 0.9158\pm 0.0039$.\foot{The proximity of the
 absolute value of this number 
to the value of the Catalan's  constant was noticed by the authors of \ci{rtt} 
but the final result for
the coefficient $\c_3$  in the original version of \ci{rtt}  was 
incorrect
due to several errors which were finally corrected  in the revised version 
 (\ci{rtt},v4).} 

Very recently the  analytic results   for the coefficients 
in the strong coupling expansion of the solution of the BES equation 
for the    cusp anomaly function \rf{f} was
 found in a remarkable paper of \ci{bkk}, 
with the first few  leading coefficients given by\foot{The relation of 
 the notation
used in 
\ci{bkk} to ours  
is: \ $\G_{\rm cusp}(g)  = \ha f(\l) , \  c_k = - { 1 \ov (4 \pi)^k} \c_k , 
\  g= {\sql \ov 4 \pi}$. We do not shift the argument of 
cusp anomaly function $\G_{\rm cusp}(g)$  by $c_1$  as was done in \ci{bkk}.}
\bea\la{tw}
\c_2&=& - \KK \ , \\ 
 \c_3&=&  - {\textstyle{ 1 \ov  32}} \big[{27}  \zeta(3) +  96  {\rm K} \ln 2   
  \big]   \  , \la{tr}\\
    \c_4 &=& - {\textstyle{ 1 \ov 16}}   \big[84\beta(4) +  81 \zeta(3)\ln 2  +
     32  {\rm K}^2 
   +144 {\rm K}(\ln 2)^2   \big] \ 
     , \la{fo}
     \\
\c_5&=& -{\textstyle{9 \ov 2048}  }
  \big[ {\textstyle {4785}} \zeta(5) +{\textstyle10572}
 \beta(4)\ln 2  +{\textstyle 4416} \zeta(3) 
  {\rm K}+{\textstyle 5184}\zeta(3)(\ln 2)^2  
  + {\textstyle 4096} {\rm K}^2\ln 2
 \big]                         
       \la{pya}  \eea
where 
\be \la{ze}
\z(k) = \sum^\infty_{n=1} { 1 \ov n^k} \ , \ \ \ \ \ \ \ \ \ \ \ 
\beta(k) = \sum^\infty_{n=0} { (-1)^n \ov (2n+1)^k} \ , \ \ \ \ \ \ 
\beta (2) = \KK   \ . \ee
The expression  for $\c_2$ \rf{tw}  thus agrees with the numerical 
value found in \ci{ben} 
and  matches precisely (the corrected version of) the 
result of the 2-loop superstring computation in \ci{rtt}. 
 
 \bs

Our aim here  is  to confirm
 the Catalan constant value of $\c_2$ in \rf{tw}
by an independent   2-loop superstring  computation.
The agreement of the results for $\c_2$ obtained in \ci{ben} and \ci{bkk} 
from the BES equation 
%
%
 with our  superstring expression   provides an important 
  test of the  BES equation  and thus of 
the underlying  asymptotic Bethe ansatz.
The significance of the result of the present paper 
 is  that it provides a highly non-trivial 
 confirmation of the proposal for the all-order strong-coupling phase \ci{bhl} 
 and its weak-coupling  continuation in \ci{bes}. 
 Indeed, while  the expressions for the tree-level \ci{afs}  and 
 the  1-loop \ci{bt,hl} 
  terms in the strong-coupling expansion   for the 
 phase where essentially put into the Bethe ansatz expression 
  from the known string theory results, the 
  higher order terms in the phase  where  conjectured in \ci{bhl} 
  using the crossing symmetry condition  \ci{jan} (which so far was not 
  directly derived from string theory). 
  The present computation demonstrates that the 2-loop term in the phase 
  suggested in \ci{bhl}  is indeed in agreement with string theory.

\bs

The  computation described below  resolves also
a technical problem related to UV regularization
present in  the original approach  of  \ci{rtt}.
The  manifest cancellation of the logarithmic UV divergences that we find here 
provides  a  direct 
demonstration   of the quantum consistency  of the \adss Green-Schwarz (GS) 
action  of \ci{mt}.  
This (together with the  earlier 1-loop results  \ci{ft1,ft2})
  removes  any doubt that this action  can be used as a basis for 
non-trivial   strong-coupling computations in the  AdS/CFT.
The agreement with the Bethe ansatz result 
provides also an implicit check of the quantum integrability of this 
$AdS_5 \times S^5$ superstring theory.

Another new result is the   suggestion 
of   a 2d Feynmann diagram  (i.e. quantum superstring) interpretation 
to the higher-order coefficients  \rf{tr}--\rf{pya}, etc.  found in \ci{bkk}.
In  our computation $f(\l)$ appears 
in the quantum 2d effective  action  of the \adss  superstring sigma model 
expanded near a particular  ``homogeneous''  string  background in $AdS_5$ 
\be \la{ga}
 \G= - \ln Z = \ha f(\l) V_2   \ .   
\ee 
 $\G$ is proportional to the  (large)  volume  factor $V_2 $.\foot{For a homogeneous
 backgrounds such as those considered in \ci{rtt} and 
 here  there is 
 no distinction between the 1-PI effective action 
 and the logarithm of  the partition function $Z$: 
  connected  but not 1-PI irreducible 2d Feynman graphs vanish.}
 This  2d QFT  interpretation  of $f(\l)$ implies  that different 
 parts of the transcendental coefficients  $\c_L$   appearing  
 in \rf{f},\rf{tw}-\rf{pya}
 can be  associated with  the contributions of different $L$-loop 
 Feynmann diagrams in the superstring sigma model. 
 
 In the 2-loop case  both the bosonic and the fermionic ``sunset''
  diagrams (Figures 1a and 1c) happen to 
 contribute  terms proportional to $\KK$  (see \ci{rtt} and below).
\begin{figure}[ht]
\centerline{\includegraphics[scale=0.5]{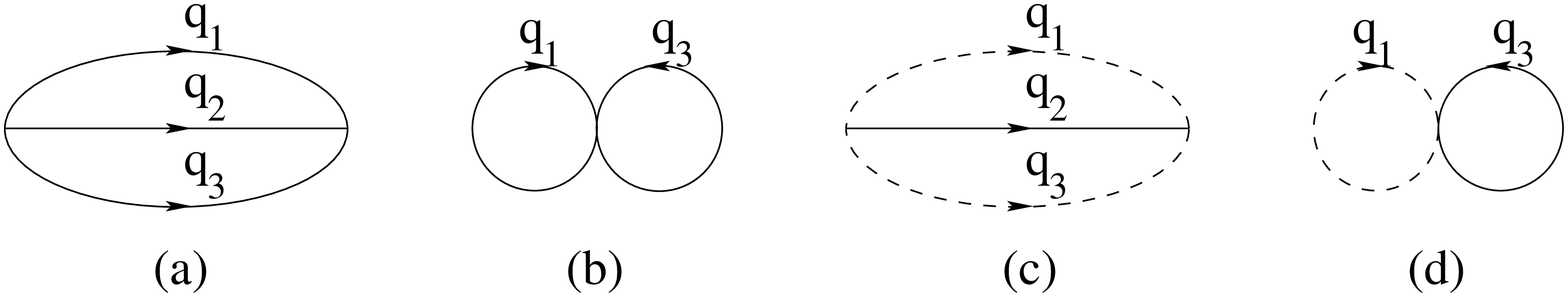}}
\caption{Two-loop diagrams (bosonic propagators are denoted by solid lines
 and fermionic ones are denoted by dashed
lines). 
\label{topthree}}\nonumber
\end{figure}
%
%
  Extending 
 our superstring computation to the 3-loop order appears to be relatively straightforward. 
 A qualitative analysis shows   that  $\zeta(3)$ term  in $\c_3$ in \rf{tr}
 should originate from diagrams in 
  Figures 2b, 2e, 2g and 2h, while the   $ {\rm K} \ln 2$ term  
   should come  from  diagrams  in 
 Figures 2c  and 2f.
 \begin{figure}[ht]
\centerline{\includegraphics[scale=0.475]{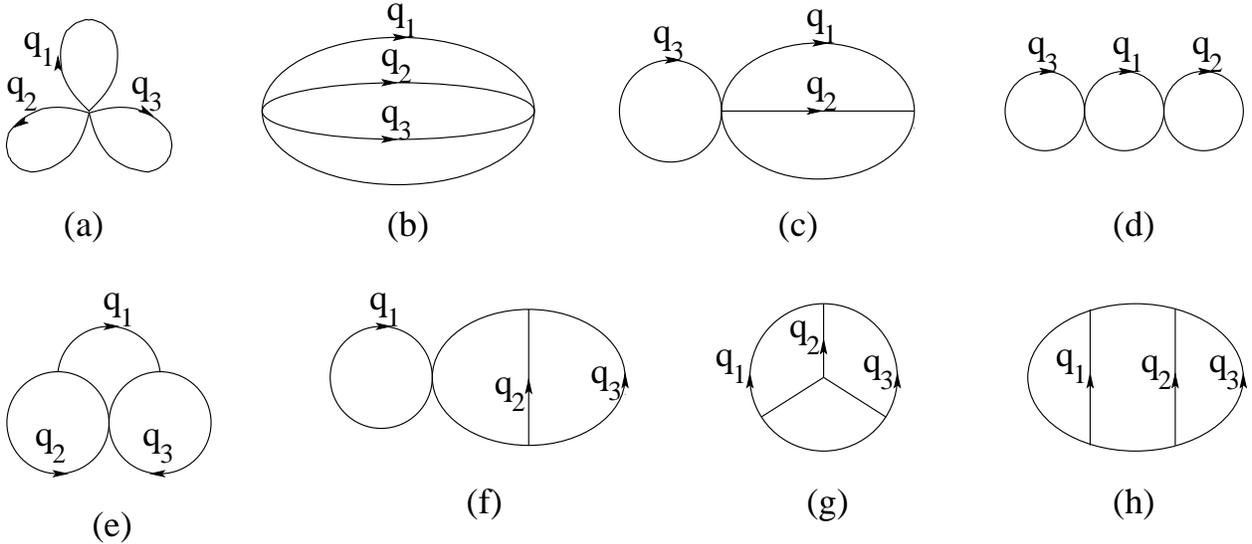}}
\caption{Topologies  bosonic  three-loop diagrams 
(diagrams  with fermionic lines have similar topology). 
\label{ndiagrams3loop}}\nonumber
\end{figure}
 In general, it is natural to conjecture that 
 the ``maximally irreducible''  terms $\zeta(2m+1)$ in   the 
 coefficients $\c_{2m+1}$ 
 and $\beta(2m)$ in  
 the  coefficients $\c_{2m}$  \ci{bkk} 
 should originate,  respectively,   from the ``maximally irreducible'' 
 odd-loop $L=2m+1$   and even-loop  $L=2m$ 
 superstring Feynman
  diagrams.\foot{This 
  should  apply starting with 2-loop order. Using this logic at the 
    1-loop order 
 one  would get   $\c_1 \sim \zeta(1)$   but this is 
 logarithmically 
 divergent; in fact,  the 1-loop divergences cancel 
 between bosons and fermions 
 and the finite remainder happens to be proportional
  to $\ln 2$ \ci{ft1,krtt}. The 
  1-loop tadpoles adjoined to
   lower-loop topologies should perhaps be interpreted in this way.
 }
 
 \bs
 
 This string world-sheet, i.e.   2d QFT  interpretation of the function 
    $f(\l)$ 
 may  help to clarify the meaning 
 of the Borel  non-summability 
 of the strong-coupling expansion 
 for $f(\l)$  as found from the BES equation 
  in  \ci{bkk}. As was  
 observed in 
   \ci{bkk}, all coefficients $\c_k$  in \rf{f} except the first 
   one are {\it negative} 
   and their values grow factorially (cf. \rf{tr}--\rf{pya}). 
   It appears that in contrast  to  sign-alternating Borel-summable series 
   usually found in  QM or QFT problems  with perturbatively stable vacuum 
   here we are dealing with an expansion near an unstable point. This is puzzling 
   since  the  rotating folded  string solution or 
   the    null cusp  
    solution of \ci{kru} we  consider below   (which are closely related \ci{krtt,am2})
   are perturbatively stable.\foot{In the conformal gauge we will be  using here
   there  is formally a ghost  fluctuation mode 
   corresponding to  the time direction in $AdS_5$ but like in  the 
    flat Minkowski space  case 
   or in the  $AdS_3$ WZW model  the underlying string theory  should be unitary: 
   the Virasoro  condition selects only physical on-shell modes.  In our 
   conformal-gauge partition function computation we are expanding 
    near a consistent 
    on-shell string  background   so the unphysical modes (a massless 
    time-like (ghost)  fluctuation 
    mode  and another  massless longitudinal mode) 
    should decouple and they 
     actually do (their trivial 1-loop contribution cancels against 
     that of the  conformal gauge ghosts).} 
   One may contemplate  the presence of 
   some  non-perturbative instability. 
   We shall further comment on this in the concluding section 4.

\bs

The rest of the paper is organized as follows.
We shall start in section 2   with setting up the  computation 
  of the  
the   cusp  anomaly function using the 
open-string (Wilson line \ci{poly,maldrey}) 
   approach 
 which is 
based  on expansion  near a  Wilson line surface with a  null cusp \ci{kru,krtt}.
As was explained in \ci{krtt,am1,am2}
it is equivalent to the closed-string approach used in \ci{gkp,ft1,ftt,rtt}.
We shall use the \adss  GS superstring action in a special $\k$-symmetry gauge 
which becomes {\it quadratic}
 in fermions  \ci{kt} after the T-duality along the 4 $AdS_5$ boundary 
  directions   in the Poincare coordinates.\foot{This   action 
was found in \ci{kt} by  starting  with  the action of \ci{mt}  written 
in a  special $\k$-symmetry gauge 
discussed in   \ci{kar}. 
An equivalent action  which  also   becomes   quadratic in fermions
after the T-duality  
  was found in  a similar   $\k$-symmetry  gauge (``S-gauge'') 
   in  Appendix C of \ci{mts}.}  
   This action was already used in \ci{krtt} 
    for the computation of the 1-loop coefficient $\c_1$ 
   in \rf{ff}. Here we shall utilize 
   its simple structure (in particular, the absence of the quartic
   fermionic terms)   to  perform  the  computation of the 2-loop coefficient 
    $\c_2$.

 \bs
 
In section 3 we shall turn to computation of quantum corrections to string
 partition function expanded near the ``null cusp''  string background. 
We shall discuss the issue of UV regularization, 
pointing  out that the structure of the superstring action involving the 
$\epsilon^{ab}$ tensor 
in the fermionic term 
prohibits 
the use of 
a direct version of the 
2d dimensional regularization.
Its use is not actually necessary since  we find that   all 
the  logarithmic 2-loop divergences  cancel out separately in
 the sums of the 
bosonic and fermionic graphs computed directly in $d=2$. 
The remaining power divergences can  then  be eliminated 
using  a  kind   of analytic
regularization which  essentially amounts to setting 
$\delta^{(2)}(0)=0$.\foot{In principle, one 
should be able to show the cancellation of all power-like divergent terms 
directly,  by carefully 
including the
contributions of all local factors (measure, $\kappa$-symmetry ghosts, 
Jacobians due to change of  fluctuation bases, etc.). Bosonic power-like 
divergences are indeed 
cancelled by the invariant  measure   contribution  \ci{rtt}. 
The same  should apply to 
the  fermionic sector:
  as was discussed in Appendices  C and D.1 in \ci{rtt}, the  
cancellation of the 2-loop power-like divergences  is required  in order  for 
the superstring 
partition function to be equal to 1  in  supersymmetric cases 
such as the flat space GS  action  expanded near a long fundamental 
string background and the \adss GS action expanded near a BMN geodesic.} 
This should be considered as a regularization prescription that 
defines the quantum \adss superstring theory in a way consistent 
with its classical symmetries, 
i.e. as a conformal quantum 2d field theory. 

 As summarised in section 3.2, the 
  resulting finite  contributions to the 2-loop coefficient in \rf{f} coming 
from the bosonic  and from the fermionic 2-loop graphs 
in Figure 1  happen  to be  the same 
as found  
in the closed-string picture computation  in  \ci{rtt}
\be 
\la{fin}
\c_2 = \c_{2B} + \c_{2F} = \KK - 2 \KK = - \KK \ ,  \ee
 so that the total result  matches the value in \rf{tw}.
 Higher-loop generalizations   are   discussed in  section 3.3.

 Section 4 contains some remarks on the problem with summability 
 of the series in \rf{f}  and also on a possible 
  generalization of 
 the present  2-loop  computation to the case of  non-zero 
 angular momentum $J$ in $S^5$.

 Some  technical details related to  the structure of the 
 fluctuation Lagrangian from section 3 are 
 given in Appendix.

\renewcommand{\theequation}{2.\arabic{equation}}
 \setcounter{equation}{0}

\section{Superstring action, classical 
 string background  and  fluctuations}

Our starting point  will be  the path integral  with the 
 euclidean version of the  $\k$-symmetry gauge fixed and T-dualized 
 \adss action  found in  
\ci{kt}. This action   is remarkably simple 
being   quadratic in  fermions ($m=0,1,2,3; \ s=4,...,9, 
\ \ z^2= z^s z^s, \ a,b=0,1$)\foot{We shall mostly  follow the notation of \ci{krtt}. We 
  choose  the  conformal gauge and 
 ignore the dilaton coupling originating from  the 2d duality transformation.
We also use Euclidean  signature on the world sheet, i.e. 
$\s^a= (\s^0,\s^1)$ ($\s^0= - i \tau)$ as  appropriate
 for the null cusp solution of
\ci{kru}; thus   there is no $i$ in front of
 the fermionic term.} 
\begin{equation}
S_E=\frac{\sql}{4\pi }\int d^2 \sigma \bigg[
\frac{1}{z^2}(\partial^a x^m \partial_a x_m+ \partial^a z^s
\partial_a z^s)
+ 4  \epsilon^{ab}  \bar{\theta}(\partial_a x^m \Gamma_m+ 
\partial_a z^s \Gamma_s) \partial_b \theta\bigg]
 \ . \la{acc} \end{equation}
Here $\theta$  is a Majorana-Weyl  
 10d spinor and $\G_A$ are standard ``flat'' 10d Dirac matrices. 
 For our present purpose the use of this  T-dual action is  a technical trick
 that allows us to reduce the number of fermionic 2-loop diagrams
  we should
  compute.\foot{
  This T-duality is a quantum symmetry when the both world sheet 
  directions are non-compact (as is the case  in our present discussion). 
  The   T-duality  maps the bosonic  \adss part of the action 
  into an  equivalent \adss  bosonic sigma model ($z \to z^{-1}$ is a  symmetry 
  transformation). Thus  
   the bosonic \adss  action has two 
  different GS superstring extensions with different fermionic parts:
   the familiar one \ci{mt} corresponding to the near-core  D3-brane 
  background where \adss  space is supported by the RR 5-form flux and the 
  T-dual one corresponding \ci{kt,krtt} to the near-core smeared  D-instanton  background  
  where  the \adss space is supported by the RR scalar and dilaton.
  For  other potential applications of this T-dual 
  action see \ci{am1,krtt}.}   
 
 It will be  useful to split the 6 coordinates $z^s\equiv z \hat
 z^s, \  \hat z^2 =1$ orthogonal to the  directions $x^m$ 
 along the boundary of $AdS_5$ as 
  ($s'= 4,...,8$) 
 \be\la{zes} 
z^{s'}\equiv z{\hat z}{}^{s'}=  z\frac{y^{s'}}{1+\frac{1}{4}y^2}~, \ \ \ \ \ \ \ 
z^9\equiv z{\hat z}{}^{9} =  z\frac{1-\frac{1}{4}y^2}{1+\frac{1}{4}y^2}  \ , 
\ \ \ \ \ \ \ 
\frac{dz^s dz^s}{z^2}=\frac{dz^2}{z^2}+\frac{dy^{s'} 
dy^{s'} }{\left(1+\frac{1}{4}y^2\right)^2}
\ee
where $y^{s'}$ parametrize $S^5$.

\subsection{``Null cusp'' solution}

The conformal-gauge form of the solution for the open string world sheet ending on 
two light-like lines forming a cusp at the boundary is \ci{kru,krtt}
\be 
&&
{\bar z}=\sqrt{2}\ e^{-  \a  \uo -  \b  \ut}~, ~
\la{ba}\\
&&{\bar x}^0=e^{-\a \uo -\b \ut}\cosh(\b \uo-\a \ut)\ , \ \ \ \ \ \ \ 
  {\bar x}^1= e^{-\a \uo-\b \ut}\sinh(\b \uo-\a \ut) \ ,   \la{bac}\\
  && ~~~~~
\a^2+\b^2=2 \ , \la{vi}
\ee
where  all other coordinates vanish ($\bar x^{2}= \bar  x^3= \bar y^{s'} =0$) 
and the last equation \rf{vi} follows from the conformal gauge condition.
The original solution 
of \ci{kru} corresponds to  $\a= \sqrt 2  ,  \  \b= 0$.

The boundary $z=0$  is reached in the limit $\s_a \to \infty$ (assuming that 
$\s_a$ run in the infinite range   and $\a,\b \geq 0$).
The induced metric is $ds^2_2= d\s_0^2 + d\s_1^2$
so that  the value of the classical action is  simply 
\be \la{kl} 
\bar S_E= { \sql  \ov 2 \pi}  V_2 \ , \ \ \ \ \ \ \ \ \  V_2 = \int d^2 \s   \ .  \ee
For cusp anomaly interpretation  it   requires a regularization as discussed in
\ci{kru,am1,krtt}. This  issue   will not be important for us here as quantum corrections to 
\rf{kl} will also scale as $V_2$  and we will be interested in the value of the 
overall coefficient $f(\l)$  in   \rf{ga}.

 The free parameters $\a,\b$  included for
generality, reflect the possibility of making $SO(2)$ rotations of the world-sheet coordinates
$\s_a$
which leave invariant the sigma model  conformal-gauge equations of motion and constraints.

Under the  2d duality  (T-duality) transformation 
$ z^{-2} \del_a x^m  \to \ep_{ab} \del^b \td x^m$
the  solution \rf{ba}-\rf{vi} is essentially mapped into itself: the duality is equivalent to 
interchanging $\s_0 \to \s_1, \ \s_1 \to - \s_0$   and inverting 
$z$ which can be implemented by changing the signs of $\a,\b$   and 
shifting $\s_a$ by constants.
This is  the reason  why,    
instead  of starting with the original \adss action 
(containing quartic fermionic terms)  and  expanding  the string path integral 
(giving the  expectation value of the corresponding Wilson loop on the 
gauge theory side \ci{poly,maldrey}) near the
 null cusp solution \ci{kru} in order to extract from it the 
cusp anomaly coefficient $f(\l)$, 
we may formally  start with the T-dual action \rf{acc}
and expand it near the equivalent null cusp solution  in  \rf{ba}-\rf{vi}. 
As was already checked   in \ci{krtt}, this procedure  leads indeed  to the same   1-loop
coefficient $\c_1$ in \rf{ff} as found  in the closed string approach (i.e.
in the energy of the closed spinning string).

To expose the fact that the $SO(2)$   2d  Euclidean rotational 
 invariance of the conformal-gauge  string  sigma model\foot{In general, the Euclidean  classical string sigma model  equations  and 
 conformal gauge constraints are  covariant under the residual
 holomorphic conformal transformations
 of $\s_1 + i \s_0$  and $\s_1 - i \s_0$.} 
  is only
spontaneously broken  
by the classical background, it is useful to write the solution \rf{ba}-\rf{vi} 
in terms of  the 2-vectors $n_{1a}, n_{2a} \  (a=0,1)$
\be\la{nn} 
n_1={ 1 \ov\sqrt{2}} (\a, \b)\ , 
~~~~~~~~
n_2={ 1 \ov\sqrt{2}} (-\b,\a) \ , 
~~~~~~~~
n_1\cdot n_1=n_2\cdot n_2=1\ , 
~~~~~~~~
n_1\cdot n_2=0~~, 
\ee
i.e. ($n \cdot \s \equiv n^a \s_a$) 
\be\la{bc} 
{\bar z}=\sqrt{2}\ e^{-\sqrt{2}\ n_1\cdot \uu} \ ,~~~~
~~~~ \ \ \ 
{\bar x}^0\pm {\bar x}^1 =e^{-\sqrt{2}\ (n_1\cdot \uu  \pm  \ n_2\cdot \uu)} \ .
\ee
In particular, in the  simple case of $\b=0$ we have 
\be \la{ch}
 \a= \sqrt 2 \ , \ \ \ \ \ \ \ \ \  \b= 0  \ , \ \ \ \ \ \ \ \ \ 
n_1= (1, 0)\ , 
~~~~~~~~n_2= (0,1) \ . 
 \ee
 Below we   will express   the string  Lagrangian for fluctuations near  the null cusp solution 
 in terms of the constant vectors $n_a$. 
 This  will help to make  the structure of the quantum contributions 
 more transparent. 
  
  \bigskip
 
 \subsection{Fluctuation Lagrangian}

 To find  the string fluctuation Lagrangian near the background \rf{bc} it is 
 useful to utilize the observation of \ci{krtt} that written in global $AdS_5$ 
 coordinates it can be related (by an $SO(2,4)$ isometry and 
 an analytic continuation) to the scaling limit \ci{ft1,ftt} of the  spinning 
 closed string solution of \ci{gkp}. The latter  background is effectively
 homogeneous,\foot{One can make this explicit  by an analytic continuation to an 
 $S^5$ solution \ci{ftt,rtt} or directly by a  special choice
  of coordinates in $AdS_5$
 as discussed in \ci{krtt,am2}.} 
 i.e. the corresponding fluctuation  Lagrangian should have  constant coefficients
 after an appropriate  choice of  basis of the 
fluctuation fields.\foot{To make  the homogeneous nature of the solution \rf{bc} explicit 
it is useful to choose a different set of coordinates in 
the Poincare patch of $AdS_5$: $ 
ds^2 = dr^2 + e^{-2r} dx^m dx_m  = 
dr^2 + (dh^m + h^m dr) (dh_m + h_m dr) 
$
where $z= e^r$ and $h^m = {x^m \over z} = e^{-r} x^m$  ($m=0,1,2,3$ and the metric has signature
 $(-,+,+,+)$).
Next,  we set  $
h^\pm  = h^0 \pm  h^1 = v e^{\pm w}$.
Then the $AdS_5$ metric above takes the form 
$ 
ds^2 = dr^2  -  (dv + v dr)^2  + v^2  dw^2 
+ (dh_i + h_i dr) (dh_i + h_i dr) 
$, so that 
the   shifts of $r,w$ are  linear isometries ($i=2,3$). 
Let us assume that the world-sheet signature is euclidean
and consider the corresponding string action in conformal gauge. 
Then simplest solution to look for is a homogeneous one
where only the two isometric coordinates are non-zero and linear:
$
v= v_0={\rm const}, \   r= k_a \sigma_a , \   w= m_a \sigma_a , \ \ \
h_i=0$
($k$ and $m$ are   constant 2-vectors). 
Note that this ansatz makes sense only for an infinite open string 
since $r$ and $w$ are non-compact coordinates.
The equations of motion are  satisfied if $k^2 = m^2$ 
and the conformal gauge constraints give (assuming the induced metric has standard flat form):
$(1 - v_0^2) k_a k_b +   v_0^2 m_a m_b =   \delta_{ab} $, i.e.
$ k^2=m^2=2, \ k_a m_a =0   ,\  v_0^2 = \ha$.
This  gives 
 $ 
z= e^{ k \cdot \sigma}  , \ \ 
x^\pm = v e^r h^\pm = {1 \ov \sqrt 2}  e^{ (k \pm  m)\cdot \sigma}, $  i.e. 
brings us back to solution \rf{bc}  after a trivial rescaling of $z$ and $x_m$ and renaming of the constant
vectors.
The fluctuation Lagrangian written in terms of $r, v, w, h_i$ 
  has only  constant coefficients  and at most quartic vertices.
  Let us mention  a generalization of the above solution to the case when 
  there is also a ``rotation''   in $S^5$ direction. The solution related to the  
  scaling limit \ci{ftt}
  of $(S,J)$ string \ci{ft1}   in the same way as described in \ci{krtt}
  has also a non-trivial angle $\varphi= \nu' \s_0$
  of $S^5$ 
  ($\nu' = i \nu, \ \ J= \sql \nu$ in Minkowski signature)  and 
  $v= v_0= { 1 \ov \sqrt 2}, \ \  r= - \kappa  \s_0 + \mu \s_1  + \ha  \ln 2 , \ \ 
  w= \kappa  \s_0 + \mu \s_1 , \ \  \kappa^2 = \mu^2 - \nu'^2  $.
  Equivalently, $z=\sqrt 2 e^{- \kappa  \s_0 + \mu \s_1}, \ \ 
  x^+ = e^{2 \mu \s_1}\ , \ \
  x^- = e^{-2 \kappa \s_0}$. 
  The conformal factor of flat induced metric is equal to 1 when $\mu=1$.
  }

 In  conformal gauge  the $AdS_5$ and $S^5$ parts of the 
 bosonic fluctuation Lagrangian are decoupled  and can be written as 
  (see  also \ci{krtt}) 
 \bea\la{bf}
&&\td L_B
 = { \sql \ov 4 \pi}  \L_B \ , \ \ \ \ \ \ \ 
  \L_B=\L_{\rm AdS_5} + \L_{\rm S^5} = \L_2 + \L_3 + \L_4  + ...    \ , \\
&&
\L_{\rm 2,AdS_5}= -(\partial\tphi)^2
+{ \frac{1}{2}}(\partial \tfpo)^2 +{\frac{1}{2}}(\partial \tfpt)^2+4
\tphi(n_2\cdot\partial\tfpo + n_1\cdot\partial\tfpt ) \cr 
&& \ \ \ \ \ \ \ \ \ \ \ \ \ 
+\ ( \partial \txx)^2+(\partial \tyy)^2 + 2 \txx^2 + 2 \tyy^2 \ , \la{bi} \\
&&
\L_{\rm 3,AdS_5}=\tphi\big[(\partial\tfpt)^2-(\partial\tfpo)^2\big] - 2
 (\txx^2+\tyy^2) (n_2\cdot\partial\tfpo -  n_1\cdot\partial\tfpt ) \ , \la{bii}
\\
&&  \L_{\rm 4,AdS_5}=
-\frac{8}{3}  \tphi^3(n_1\cdot\partial\tfpt + n_2\cdot \partial\tfpo)
+(\txx^2+\tyy^2)\,\L_{\rm 2,AdS_5}-2\txx\tyy\partial\txx \partial\tyy \ , 
\la{by} \\
&&\L_{\rm 2,S^5}=  \partial y^{s'}\partial y^{s'} \ , \ \ \ \ \ \ 
\L_{\rm 3,S^5}= 0 \ , \ \ \ \   \      
\L_{\rm 4,S^5}= -\frac{1}{2} y^2\, \partial y^{s'}\partial y^{s'}  \ . \la{sss}
\eea
Here $\partial$ stands for $\partial_a$ and $n \cdot \partial = n_a \partial_a$, etc.
The background dependence is represented by 
the constant 2-vectors $n_1$ and $n_2$ in \rf{nn}. 
The fields $\tphi, \tfpo, \tfpt, \txx, \tyy$ are fluctuations 
in the five  $AdS_5$ directions,\foot{The fields   $\txx, \tyy$ 
are related to fluctuations of $x^2,x^3$ in \rf{acc} 
that  are zero in the  solution \rf{bc}.}
 while $y^{s'}$  are $S^5$   coordinates from \rf{zes} that have
zero background values.
The massless time-like (ghost) fluctuation $\tphi$ should eventually decouple 
together  with another massless longitudinal mode
(their trivial  1-loop contribution  cancels against the decoupled
 conformal gauge ghost contribution).
 
 The explicit relation between $\tphi, \tfpo, \tfpt, \txx, \tyy$
 and the fluctuations of the original  Poincare  coordinates $z, x^m$ is given in Appendix A.
 There we also  present  the  resulting bosonic propagator
 which is non-diagonal in the $\tphi,\tfpo,\tfpt$ directions.

\bigskip

Finding a  convenient (constant coefficient)  form  
for  the quadratic fermionic  term as well as
 for the fermion-boson  coupling terms 
following from  the action \rf{acc} 
requires us to perform a nontrivial 
rotation of fermions.
 This can be done in two steps.
First,  we note that 
 the world sheet position dependence in the terms involving the
 coordinates transverse to the boundary directions 
  arises entirely  from the overall factor
  of $z$ in  $z^s = z \hat z^s(y)$ (on the solution \rf{bc} we have $\bar y^{s'}=0,\ \bar {\hat
  z}^{4,...,8}=0, \ \bar {\hat z}^9=1$). 
  Redefining $\theta\mapsto \theta/\sqrt{\bar z}$ and making use of the
   identity ${\bar \theta} \Gamma^A\theta=0$ leads to the following 
   expression for the fermionic term in the  square brackets in 
   \rf{acc} 
\be\la{fi}
\L_F&=&
4\ \epsilon^{ab} \ {\bar\theta} \bigg[ {\partial_a x^m \ov {\bar z} }  \Gamma_m \ 
+ \  
\big(
{\partial_a z \ov {\bar z} } {\hat z}^s+  {z\ov {\bar z} } \partial_a {\hat z}^s
\big) \Gamma_s   \bigg]  \partial_b\theta   \ . 
\ee
Since $\bar z$  in \rf{bc}  is exponential in $\sigma_a$  
 the terms with  ${\hat z}^s$ in \rf{fi} will now  have constant coefficients
 once expanded near the solution.
A second local redefinition  of $\theta$  is needed in order 
to take into account that 
 $x^0$ and $x^1$ 
have nontrivial backgrounds in \rf{bc}. 
In general, the background value ${\bar N}$ of 
\be\la{nu}
{N}_a^u \equiv \frac{\partial_a {x}^u}{\bar z}~, \ \ \ \ \ \ \ \ \ \ \ ~~~u=0,1; \ \ a=0,1
\ 
\ee
is not an  $SO(1,1)$ rotation  matrix:
\be\la{bu0}
{\bar N}_a^u{\bar N}_b^v\eta_{uv }= n_{2a}n_{2b}-n_{1a} n_{1b} \ . 
\ee
It is nevertheless possible (though somewhat complicated) 
to find an $SO(1,1)$ rotation of fermions that removes the position dependence
from their action. For simplicity, it is sufficient to  consider the 
case of $\beta=0$ in \rf{ch}. Then we get 
\be\la{bu}
{\bar N}_a^u{\bar N}_b^v\eta_{uv }=\eta_{ab} \ ,  
\ee
and thus the required  $\sigma_a$ dependent 
rotation of $\theta$ is 
\be
\theta \ \ \mapsto \ \  \big[  \cosh ({1 \ov \sq}\ n_2\cdot \s)  +  \sinh ({1 \ov \sq}\  n_2\cdot \s ) \ 
 \Gamma^0\Gamma^1\big] \ \theta  \ . 
\la{ro}
\ee
Moreover, it  turns out that the matrix 
\be\la{baa}
{\cal N}_{ab}=N_a^u {\bar N}_b^v \eta_{uv}
\ee
expanded near  the classical solution  has only terms with  
 constant coefficients  in front of the  
 bosonic fluctuations  
 (its expression to leading order in bosonic fluctuations is
  given in Appendix A).\footnote{In the general case  of 
 $\beta\ne 0$  when  ${\bar N}_a^u{\bar N}_b^v\eta_{uv }$ is still 
 a constant 
 off-diagonal matrix
  the expansion
 of ${N}_a^u{\bar N}_b^v\eta_{uv }$ around the classical solution 
 has again the constant coefficients.}

Taking into account the effect of the rotation \rf{ro} 
and making further use of the identity ${\bar \theta} \Gamma^A\theta=0$, 
we finally find for the  fermionic part of the fluctuation Lagrangian 
($a,b=0,1; \ i,j=2,3$;\ $s, t=4,5,\dots, 9$): 
%
%
\be
\L_F&=& 4\epsilon^{ab}\ {\bar\theta}\Big[ -{\cal N}_{ac} \Gamma^c
+ \
   \frac{\partial_a x^{i}}{\bar z} \Gamma_i  +
\big(
\frac{\partial_a z}{\bar z} {\hat z}^s+\frac{z}{\bar z} \partial_a 
{\hat z}^s\big)    \Gamma_s \Big]\partial_b\theta
\cr
&&-\ 2 \sqrt 2\
\epsilon^{ab} \ n_{2b} \ 
{\bar\theta}
\Big[ \frac{\partial_a x^{i}}{\bar z} \Gamma_i  
+ 
\big(
\frac{\partial_a z}{\bar z} {\hat z}^s+\frac{z}{\bar z} \partial_a {\hat z}^s
\big) \Gamma_{s} \Big]\Gamma^0\Gamma^1\theta \ . 
\la{fii}
\ee
The second line  appears due to the rotation \rf{ro}.
The bosonic fields here 
 can be expanded in fluctuations $\tphi, \tfpo, \tfpt, \txx, \tyy$
 and $y^{s'}$  (using the relations in Appendix A) 
  leading to  fermion-fermion-boson 
 and fermion-fermion-boson-boson quantum vertices needed to compute the 
 2-loop diagrams in Figure 1. 

The  quadratic  term in   \rf{fii}   determining a 
non-degenerate fermionic propagator  can be written as 
\be\la{pr}
\L_{2F}= 2\sqrt{2}\ \epsilon^{ab} \
{\bar \theta}\Big[ (-\Gamma^0+\sqrt{2}\Gamma^9)n_{1a} \partial_b -\Gamma^1 n_{2a}\partial_b
 + 
  n_{1a} n_{2b}  
    \Gamma^{019}  \Big]\theta \ . 
\ee
The  $\Gamma^{019}$ term  produces a non-zero mass  (equal to 1) 
for the  8  independent fermionic fluctuations  (see also  \ci{krtt}).

\renewcommand{\theequation}{3.\arabic{equation}}
 \setcounter{equation}{0}

\section{Quantum corrections}

Let us  now turn to the computation of quantum loop corrections
 to the effective action  as defined by path integral 
 based on the action given by the sum of the bosonic \rf{bf} and the fermionic \rf{fii} parts.

\subsection{One-loop contribution}
Let us  start with  
 reviewing the 1-loop result \ci{ft1,krtt}. 
From the quadratic part of the above fluctuation Lagrangian 
it is straightforward to recover the mass 
spectrum and the 1-loop value of the effective action \rf{ga} 
and thus  the 1-loop coefficient in the  cusp anomaly function. 

Extracting the bosonic  kinetic
operator from \rf{bf} and computing its determinant in 2d momentum representation 
(the propagator $K^{-1}_B(q)$ is given in Appendix A) 
we find
\be \la{dee}
\det K_B(q)
= -2^{8} \,(q^2)^7\,(q^2+2 )^2(q^2+4) \ . 
\ee
This means  that the bosonic spectrum contains seven massless scalars, two scalars with
mass $\sqrt 2$ and one scalar with  mass  $2$.

Performing a similar computation of the  fermionic spectrum 
from the determinant of the fermionic kinetic operator in \rf{pr}
we get ($n \times q\equiv \epsilon^{ab} n_a q_b$; see \rf{ch}) 
\be\la{fg}
\det K_F(q)= \big[(n_1\times n_2)^2 +  (n_1\times q)^2+ (n_2\times q)^2\big]^8=2^{16}(q^2+1)^8\ ,  
\ee
implying that the spectrum contains eight fermions with  mass $1$.

This coincides with
 the spectrum of fluctuations
around the folded spinning string \ci{ft1}, as was already  discussed  in \ci{krtt}.
Taking into account that the conformal-gauge  ghost contribution 
 cancels the contribution of the 
two bosonic massless modes,  the  1-loop  effective action is
found to be given by the same expression as in \ci{ft1,ftt,krtt}
\be
\Gamma_1  &=& \frac{1}{2} V_2 \int{ d^2 q \over(2 \pi)^2}
\bigg[  \ln (q^2+4)  + 2 \ln (q^2+2) + 5 \ln q^2 - 8  \ln (q^2+1)\bigg]
=- \frac{3\ln 2}{2\pi}  V_2~. \la{onn}
\ee
This leads (using  \rf{ga})  to the value of $\c_1$ in \rf{ff}.\foot{Introducing a UV cutoff
in the momentum integral  in \rf{onn} 
one finds  that the finite part proportional to 
 $-3 \ln 2$  comes only from the bosonic mode contribution  while the
role of the fermion contribution is to cancel the bosonic UV divergence.}

\subsection{Two-loop contribution}

The 1-loop result \rf{onn} is manifestly finite: the 
logarithmic UV divergences cancel between the  bosonic   and the   fermionic terms. 
 As 
was discussed  in detail in \ci{rtt}, the 
 issue  of potential higher-loop  UV divergences  in 
 Green-Schwarz 
action expanded near a particular string
 background is subtle, due in particular  to the  lack of manifest
power counting renormalizability.\foot{As for the  2d IR divergences, they  cancel 
in on-shell effective action as expected on general grounds \ci{rtt}.}

To get rid of power divergences in \ci{rtt} we attempted to use 
dimensional regularization (as is common in the treatment  of 2d sigma models).
Continuing the \adss  GS action to $d=2-2\eps$ dimensions appears, however,
 to be
inconsistent as this 
spoils its  classical $\kappa$ symmetry.\foot{Ideally, the regulator should be introduced before gauge fixing
so  that it preserves all local invariances (and as many of the global 
invariances as possible). The presence of the Levi-Civita tensor (WZ) 
 term in 
the \adss   GS action 
and   the related 2d self-duality  property  of the $\kappa$-symmetry parameters 
makes 
dimensional continuation problematic.}
 While the regularization procedure
used in \ci{rtt}  made it  possible to  find the non-trivial finite part 
of the 2-loop effective
action and reproduce
 the value \rf{tw} of $\c_2$,\foot{The  value  of $\c_2$ 
 in the original version of \ci{rtt}  was incorrect: 
 (i) the cancellation of the second transcendental constant $\td \KK$ 
in the bosonic contribution was overlooked; (ii)  the normalization of 
the fermionic  contribution was
off by factor of 2; (iii) the computation was done 
 for an $S^5$ background related to the 
relevant $AdS_5$ rotating string background by an analytic 
continuation \ci{bfst,ftt} that also inverts the sign of the  string
tension, so that the result of the 2-loop $S^5$ computation 
should be taken at the end with an opposite
sign. These errors were corrected in the revised version (v4) of \ci{rtt}.}
  it did not 
allow us  to  check  the
expected cancellation of the logarithmic divergences.

Here we  resolve this problem. A consistent computational procedure appears to be 
as follows. One should first not use any explicit regularization 
and rearrange the   momentum integrals  (directly in $d=2$) 
to extract 
all potential logarithmically divergent contributions.\foot{The procedure of 
required rearranging
of the momentum integrals 
by reducing the power of momenta in
 the numerators was described in detail in \ci{rtt}.}
 Remarkably, by  direct computation    of the 2-loop graphs in Figure 1 
 starting with the action \rf{acc},\rf{bf},\rf{fii} 
we have found  that the thus extracted $\ln \Lambda$ and $\ln \Lambda^2$ 
 logaritmic divergences {\it  cancel} 
{\it separately}  in the sum of purely bosonic  graphs (Fig. 1a, 1b) 
and  the sum of graphs with fermionic propagators (Fig. 1c, 1d).
The remaining  power divergent terms  can then be regularized away 
by a kind of   analytic regularization prescription.
 In fact, they  should  cancel 
against the invariant measure and $\kappa$-symmetry  ghost contributions in a
systematic treatment that takes into account all local $\delta^{(2)}(0)$ 
contributions.

A variant  of such  regularization procedure  
is  a version of ``dimensional regularization'' that was found in \ci{rtt}
to  preserve the BPS nature ($Z=1$) 
of the expansion near  the  BMN point-like string at 2-loops
(see Appendix C in \ci{rtt}). It assumes that the use of 
 all algebraic manipulations with  momentum integrals 
 as well as of   symmetric integration identities
is done  strictly in  two dimensions. 
The resulting  2d Lorentz {\it covariant} integrals are then 
 continued to $d=2-2\epsilon$ as a way to get rid of power divergences.
It turns out that  the simple poles in $1\ov \epsilon$  then  cancel 
at the same time as the double poles,  and that happens 
separately for the bosonic and the fermionic
 contributions. 

This prescription  amounts to a consistent definition 
(respecting   all relevant symmetries of the classical action) 
of the \adss  string theory  as a 2d quantum conformal theory. 
As we  find   below, the resulting 
 2-loop effective action  is then finite  and 
   reproduces  the value in  \rf{tw}.

\bigskip

Before turning to the summary of our 2-loop results let  us comment 
some more on the cancellation of the logarithmic UV  divergences. 
As was pointed out above,  while  at the 1-loop order the logarithmic 
UV divergences
were cancelling {\it between}
 the bosonic and the fermionic contributions 
the 2-loop 
cancellation pattern is different: 
the  logarithmic UV  divergences cancel separately in  the bosonic and fermionic
graph  contributions.\foot{ 
This may  look  surprising  given 
the well-known  expression for the  2-loop $\beta$-function
for a generic bosonic sigma model found in dimensional regularization with the minimal
subtraction scheme \ci{fri}, 
$\beta_{\m\n} = R_{\m\n} + { \a' \ov 2} R_{\m \l\r\s} R_\n^{\  \l\r\s} + ...$.
As was pointed out in  \ci{ts}, the two-loop term here 
is invariant under local redefinitions of the coupling, i.e. of the metric $G_{\m\n}$
(this is true in general for $G_{\m\n} \to G_{\m\n} +  c_1 R_{\m\n} +  c_2 R G_{\m\n} + ... $
as one  can readily check).  Thus the scheme we are using here 
is ``non-standard''  from the bosonic sigma-model point of view.
It is, however,   meaningful  in the  context of the full   UV finite 
superstring  sigma model. Note that the 2-loop beta-function  also 
   vanishes  in the ${\cal N}=1$ 2d supersymmetric sigma model \ci{frid}
   where again the use of our scheme would make sense.
}
Moreover, for the homogeneous spaces like $AdS_5$  and $S^5$ 
there is no ${ 1 \ov \eps^2} \sim  \ln^2 \Lambda $ 2-loop UV divergences
 (which are in 
 general proportional to covariant derivatives of $R_{\m\n}$). 

In fact, starting formally with such a
sigma model defined in $d$ dimensions  one finds \ci{fri,rtt}
that the potentially divergent contribution is proportional to $d-2=-2\eps$
times the square of the tadpole integral 
$I[m]= \int { d^d q \ov (2\pi)^d} { 1 \ov q^2 + m^2}$  (so that $1\ov \eps^2$ pole
cancels out). Then if one uses a scheme in which one 
  first  combines the contributions  of  momentum integrals directly in $d=2$
  then all logarithmic divergences  cancel out. 
  In this  natural regularization prescription 
 the bosonic \adss  sigma model defined directly in $d=2$ 
 is manifestly 2-loop finite.\foot{The covariant local measure contribution cancels 
also  power divergences \ci{rtt}.}  A non-trivial check of the 
quantum consistency  of 
the \adss action is that the same applies 
 separately also to the fermionic 
graph contribution. This is indeed what we have found by direct calculation
starting with the action \rf{acc}.

\bigskip 

Given the mass spectrum  of bosonic and fermionic fluctuations described above
one may anticipate which  momentum  integrals may in principle  appear in 
the  the 2-loop effective action given 
by the sum of graphs in Figure 1.
 The   values of masses are $0, \sqrt 2, 2$ (bosonic) and $1$ (fermionic) 
  but not all combinations of masses in the propagators
 actually  happen to appear  in the final result. 
Let us define 
\be\la{ii}
I[m_1,m_2,m_3]&=&\int \frac{d^2 q_1d^2 q_2d^2 q_3}{(2\pi)^{4}}\ 
\frac{\delta^{(2)}(q_1+q_2+q_3)}{(q_1^2+m_1^2)(q_2^2+m_2^2)(q_3^2+m_3^2)} \ , 
\\
I[m_1,m_2]&=&\int \frac{d^2 q_1d^2 q_2d^2 q_3}{(2\pi)^{4}}
\frac{\delta^{(2)}(q_1+q_2+q_3)}{(q_1^2+m_1^2)(q_2^2+m_2^2)} 
= I[m_1] I[m_2] \ ,   \la{iii} \\
I[m]&=&\int { d^2 q \ov (2\pi)^2} { 1 \ov q^2 + m^2} \ . \la{i}
\ee
The  integral  \rf{ii} is  UV-finite while \rf{iii} exhibits 
$\ln^2 \Lambda$  and $\ln \Lambda$  UV 
divergences.
The  momentum integrals that appear  in the direct computation of 2-loop graphs 
starting with the action \rf{bf},\rf{fii}  can be expressed in terms of the 
sum of integrals of the  above type plus the 
power divergent contributions proportional to the square of 
 $I_0= \int { d^2 q \ov (2\pi)^2}$ 
and  to $I_0 I[m]$; these we set to zero by an analytic (e.g. dimensional)
regularization \ci{leib}.

 The explicit calculation
   has shown  that only two 
special cases of the finite integral  \rf{ii}
 remain in the final answer.\foot{All possible combinations of masses occur 
  at   the intermediate 
steps. As in \ci{rtt} the  calculations 
  were done using {\tt Mathematica}-based computer program.}
 They are \ci{rtt}
\be\la{ki}
I[\sqrt{2},\sqrt{2},2]=\frac{\rm K}{(4\pi)^2} \ , 
~~~~~~~~
I[1,1,\sqrt{2}]=\frac{2{\rm K}}{(4\pi)^2}\ , 
~~~~~~~~
{\rm K}\equiv \sum_{k=0}^\infty \frac{(-1)^k}{(2k+1)^2} \ . 
\ee
In general, the contribution of bosonic graphs to the 2-loop 
effective action in  a theory with cubic and quartic vertices is
\be\la{bop}
\Gamma_{2B}=\frac{4\pi}{\sqrt{\lambda}}\int d^2 \uu \Big(-\frac{1}{12} A_{3B}
+\frac{1}{8}A_{4B}\Big) \ , 
\ee
where $A_{3B}$ and $A_{4B}$ are, respectively the contributions of the 
graphs with  topologies shown in Figure 1a and 1b.
In the case of the bosonic $AdS_5\times S^5$
sigma model computed according to the regularization prescription described above 
they  turn out to be
\be\la{bep}
A_{3B}&=&  12 \ ( I[\sqrt{2},\sqrt{2}]+ \ 4\ I[2,2])  - 
\ 24\ I[\sqrt{2},\sqrt{2},2] \ , 
\\
\la{bip}
A_{4B}&=&     8\  ( I[\sqrt{2},\sqrt{2}]+ \ 4\ I[2,2] )  \ . 
\ee
Here  $I[\sqrt{2},\sqrt{2}]$  and $I[2,2]$ 
are the $\ln^2\Lambda$  and $\ln\Lambda $ UV  divergent integrals \rf{iii}. 
Combining them in \rf{bop} 
one finds  that they 
 {\it cancel } 
 leaving us   with a finite result proportional to the Catalan's 
 constant (see \rf{ki}) 
 \be\la{fins}
 \Gamma_{2B}=\frac{4\pi}{\sqrt{\lambda}}  \frac{2\rm K}{(4\pi)^2}  V_2  \  .\ee 
 Similarly, the contribution of   2-loop  graphs with fermion propagators 
  in  a theory with fermion-fermion-boson   and 
  fermion-fermion-boson-boson   couplings is in general 
\be\la{gh}
\Gamma_{2F}&=&\frac{4\pi}{\sqrt{\lambda}}\int d^2 \uu \Big( \frac{1}{16} A_{3F} +
\frac{1}{8} A_{4F}\Big) \ , 
\ee
where $ A_{3F}$ and  $ A_{4F} $ are produced, respectively, 
by graphs with topologies       in Figure 1c and 1d.
The explicit calculation yields finite results
\be\la{jk}
A_{3F}&=&- 32\  I[1,1,\sqrt 2 ]  \ , \ \ \ \ \ \ \ \ \ \   A_{4F}=0 \ , 
\ee
so that 
 \be\la{fans}
 \Gamma_{2F}=-\frac{4\pi}{\sqrt{\lambda}}  \frac{4\rm K}{(4\pi)^2}  V_2  \  .\ee 
As was already mentioned above, 
the cancellation of divergences in the sum of 
graphs with fermion propagators 
represents a strong consistency test of 
the quantum GS action. 
The absence  of UV divergences {\it separately}
 in graphs in  Figures 1c and 1d potentially suggests 
 the existence of additional non-manifest 
 symmetries in the fermionic action that 
 are preserved by the $\kappa$-symmetry gauge as well as by our 
  regularization prescription.

Combining the bosonic \rf{fins}  and the fermionc \rf{fans}
contributions, we  finish with 
\be \la{fgh}
&&\G_2  =  \Gamma_{2B}+ \Gamma_{2F } = {\c_2  \ov 2 \pi \sql}   V_2  \ , \\ \la{hk}   
&&\c_2= \c_{2B} + \c_{2F} = \KK - 2 \KK = - \KK  \ . 
\ee

Similar results  for the  finite parts of  $\Gamma_{2B}$  and $\Gamma_{2F}$
were  found also  in the  independent computation for a  closely related 
$S^5 $  background  in \ci{rtt}. The overall sign of the result for $\G_2$ 
was,  however,  opposite. This is  consistent with the equivalence of the two
$AdS_5$ and $S^5$ 
 solutions
 since the 
analytic continuation  relating them   implies that one  should 
also change  the sign of the string tension $\sql  \to  - \sql $, 
i.e.  reverse the sign of all even-loop terms in the effective action.
Formally, this  reverses  the  sign of the coefficient $\c_2$ in $\G_2$, 
leading  again to the result in \rf{hk}.

\subsection{Higher-loop contributions
 }

Going to higher, e.g. 3-loop, order is,  in principle,  straightforward.
We again expect that  all logarithmic divergences will cancel directly in $d=2$ 
while power divergences can be unambiguously separated and 
regularized away. 

Based on the spectrum of fluctuations and the form  of the 
propagators and vertices in the string  fluctuation action 
 it is relatively straightforward to determine the general
structure of the finite higher loop contributions to the
effective action and thus to the strong-coupling 
expansion of  cusp anomaly function in \rf{f},\rf{ga}
as predicted by the string inverse tension expansion.

On dimensional grounds, the finite contribution 
to the effective action or cusp anomaly comes from momentum integrals 
of mass dimension $-2$ (cf. \rf{onn},\rf{ii}). Most vertices in the action 
(\ref{acc}) contain derivatives; employing 
  partial fractioning and 2d Lorentz invariance 
these derivatives 
may be used to cancel some of the propagators. Since many  of the 
fluctuation fields in the theory are massive, 
this leaves behind terms with uncanceled 
propagators and with the momenta  in the  numerators 
replaced by the  mass  values. Thus, 
the $L$-loop contribution to the effective action can be expressed in 
terms of {\it scalar}  vacuum integrals whose topology is that of the 
initial Feynman diagrams as well as that of the ``daughter'' diagrams 
obtained by collapsing some of the propagators.

At each loop order there exists a new set of ``maximally irreducible'' topologies
(see, e.g.,  Figures 1a and 1c at $L=2$ and Figures 2b, 2e,2g and 2h
 at $L=3$). At $L$ loops these topologies
  contain at most $(L+1)$-point vertices.
All other topologies that can appear at an 
 $L$-loop order are inherited from
 lower loop orders by simply adjoining
  lower loop graphs  in such  a way that the
  total number of loops is $L$ (see Figures 1b,
   1d and Figures 
   2a,2c, 2d, and 2f).\footnote{Since
  the tadpole integral \rf{i}  is logarithmically 
  divergent and assuming that the  finiteness of the partition function 
  persists to all loop 
  orders, such graphs will either not appear at all or they will involve  
  both the bosonic and the fermionic propagators 
   in such a way  that the sum of all of them 
    is finite.}
    
The number of irreducible sigma model diagrams grows
factorially  with $L$  (there are more graphs  than, say in  $\phi^4$  theory). 
A type of graph that potentially occurs at each loop order
 is shown in Figure 3. It
\begin{figure}[ht]
\centerline{\includegraphics[scale=0.5]{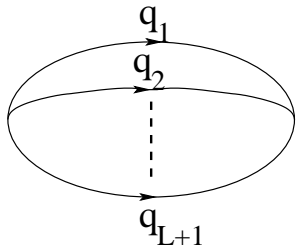}}
\caption{$L$-loop ``maximally irreducible''  sunset-type graph 
\label{topL}}\nonumber
\end{figure}
may contain various combinations of propagators  with 
mass values ($1,\sqrt 2, 2$)  from the spectrum of our  theory  (massless 
propagators  should  not appear at the end as all IR divergences should 
cancel out). 
 On general grounds,   most of  such integrals 
 may   contribute to the the effective action;\foot{Some of them may be ruled out by taking into account
 the structure of possible vertices in the superstring  action (e.g., $I[1,1,1]$
 with 3 fermionic masses is obviously not allowed).}
 explicit calculations are then necessary to fix their coefficients.
 Some of these coefficients  may,  in fact,  vanish, 
 as was, for example, 
  the case with  
 $I[2,2,2]$ in the  bosonic part  \rf{bop} and  with $I[1,1,2]$  
 in the fermionic part of the    2-loop result (see also \ci{rtt}).  

\bs

To examine which  new transcendental numbers 
 may possibly  appear as coefficients in $\G$  at the 3-loop order let
us consider a generalization of the integral \rf{ii} 
\be
I[m_1,m_2,m_3,m_4]=\int\frac{d^2q_1d^2q_2d^2q_3d^3q_4}{(2\pi)^8}
\frac{\delta^{(2)}(q_1+q_2+q_3+q_4)}{(q_1^2+m_1^2)(q_2^2+m_2^2)(q_3^2+m_3^2)
(q_4^2+m_4^2)} \ , 
\ee
which corresponds to a graph  of  ``maximally irreducible'' 
 topology shown in Figure 3 with  $L=3$ (i.e. to Figure 2b). 
It is not easy 
to compute this integral for arbitrary values of the masses $m_k$. Using 
 the values of $m_k$ that
appear in our  spectrum of fluctuations and making 
 a  simplifying assumption 
  that the   four masses split into two equal pairs,
  we find:\footnote{The first integral here is related to the one 
    found   in \ci{br}. 
 It is interesting to note that integrals with irrational mass ratios do
   not have definite transcendentality.}
\be
I[\sqrt{2},\sqrt{2},\sqrt{2},\sqrt{2}]&=& 2 I[{2},{2},{2},{2}] =
 \frac{7\ \zeta(3)}{2 (4\pi)^4} \ , \la{zzzz}
\\
I[\sqrt{2},2,\sqrt{2},2]&=& \ha I[1,\sqrt{2},1,\sqrt{2}]=
   \frac{1}{4\sqrt{2}(4\pi)^4}
\Big[\ln(3+2\sqrt{2})\ (\ln 2)^2             \cr
&+&4\sqrt{2}\big[\Li_2({{\textstyle {1 \ov \sqrt 2}}})
-\Li_2(  -{\textstyle {1 \ov \sqrt 2}})\big]
+8\big[\Li_3({\textstyle {1 \ov \sqrt2 }})-\Li_3(-{{\textstyle {1 \ov \sqrt 2}}})\big]\Big]  
 \ , \la{jjj} \\ 
I[1,2,1,2]&=&\frac{1}{(4\pi)^4}\Big[
\ln 3\  (\ln 2)^2 \cr
 &+&2\ln 2\, \big[\Li_2 (\ha )-\Li_2(-\ha )\big]  +2\big[\Li_3(\ha )-\Li_3(-\ha )\big]
\Big] \  . 
\la{ppp}
\ee
As follows from \rf{zzzz}, 
 the  $\zeta(3)$  coefficient makes a natural appearance at the 3-loop order
in the superstring sigma model partition
 function.\foot{As is well known, it also appears
 in the 4-loop sigma model 
$\beta$-function (computed in a mininimal subtraction scheme)
 in the case of a 
  generic target space metric \ci{zan,hlo}.}
 It is therefore very likely  that it will be present 
in the expression for  the 3-loop coefficient $\c_3$  
in  the cusp anomaly function.
This observation suggests 
 a superstring (2d Feynman-diagrammatic)  interpretation to  the value of the 
coefficient $\c_3$ in \rf{tr}  found from the BES equation in  \cite{bkk}.

Integrals for higher loop  maximally irreducible  graphs in Figure 3 
 are harder to evaluate. Their general expression  
 written in 2d coordinate space is 
\be
I[m_1,m_2,\dots,m_{L+1}]=
\int d^2 x \prod_{i=1}^{L+1}
 \Big[{1 \ov 2 \pi }   K_0(m_i|x|)\Big] \ , 
\ee
where $K_0$ is the Bessel function (${ 1 \ov 2 \pi} K_0$ is 
the  2d massive scalar propagator). 
%
%
It  would be interesting to relate their values\foot{Such sunset 
diagram integrals in different dimensions
were extensively  discussed in the literature
(see \ci{piv} and references therein)
 and can be found numerically, 
but their analytic form is  apparently not known beyond few simple examples.} 
to the constants  appearing in the expressions  
 for higher-order $\c_L$ coefficients  found in \ci{bkk}:
  the odd-loop coefficient should start with zeta function 
 $\zeta(L)$  and 
 the even-loop  one 
  should start  with the Dirichlet beta function 
 $\beta(L)$ (cf. \rf{tw}-\rf{pya}).\footnote{By writing a generating 
 function for the equal mass sunset diagrams
$
F(t)=\int d^2 x\ K_0(|x|)\ e^{-tK_0(|x|)}
$
and approximating the Bessel function in the exponent as $K_0(|x|)\sim -\ln |x|$
one may see that at $L$ loops we expect a $\zeta(L)$-type transcendentality. 
We thank A. Pivovarov for this comment.} \foot{In this
 connection let us mention the following useful relations:
 $
\beta(n)= {(-1)^n\ov 4^n (n-1)!}  [\psi_{n-1} ({1\ov 4})-\psi_{n-1} ( {3\ov 4}    )]
$, \ 
$
\beta(2n) \propto \Li_{2n-1}({1\ov 4})-\Li_{2n-1}(  {3\ov 4}   )$, \ 
$
\beta(2n-1)\propto\pi^{2n-1}
$,  where $ \psi_{n-1}(x)$   is the $n$-th derivative of $ \ln \Gamma(x)$.
They express the Dirichlet beta function in terms of 
quantities that naturally appear in loop integrals. 
} 

An underlying reason for this relation may be that 
the coordinate space  representation 
of the 2d diagrams 
is given by  integrals  of products of the Bessel function $K_0$ and its
derivatives  while the  integrals of Bessel functions appear also in the 
BES  equation \ci{bes} and its  strong-coupling solution in \ci{bkk}.

\renewcommand{\theequation}{4.\arabic{equation}}
 \setcounter{equation}{0}
\section{Concluding remarks}

One  consequence  of   the strong-coupling solution of the BES 
equation  found in    \ci{bkk} was  that the  coefficients 
$\c_1,\c_2, ...$   are all negative and grow factorially. The 
series in  \rf{f} is then  not Borel summable,   i.e. its 
summation is ambiguous and  this might  be suggesting adding 
to  \rf{f} exponentially small terms $\sim e^{ - k \sql}$ for some positive $k$. 

By formally changing  the sign of $\sql$ 
\be \la{cha}
\sql\ \  \to \ \  - \sql   \ee
one finds that \rf{f} becomes 
a sign-alternating and thus Borel summable series. 
This is  puzzling 
 since the {\it weak-coupling}  expansion 
 $f(\l)= b_1 \l + b_2 \l^2 + ...$ which is also described by the BES equation 
(and which has  finite radius of convergence) is formally 
invariant under the sign change \rf{cha} and thus is ``not  aware'' 
of the  problem with summation of the 
strong-coupling expansion.

The string theory interpretation of $f(\l)$ as a coefficient 
in  the  partition function expanded near a perturbatively stable 
string solution  would  also  suggest a standard asymptotic 
but Borel-summable  expansion in $1 \ov \sql$. 
However, the  string theory result for $\c_2$ in $f(\l)$  
found  in \ci{rtt}
and here  reproduces the negative sign of $\c_2$ in \rf{tw}  and thus 
appears to support the conclusion of \ci{bkk} about the lack of 
Borel-summability  of the strong coupling expansion of 
$f(\l)$.\foot{As was already mentioned above, the direct result of the computation 
in  \ci{rtt} was actually  the  opposite sign for $\c_2$. However, this computation 
 was done 
for a complex (but perturbatively stable)  $S^5$ solution 
 related to the scaling limit \ci{ftt} of the 
spinning string solution in $AdS_5$  by a formal complex automorphism
of the \adss string action   which is an equivalence transformation provided one also
inverts the sign of the string tension, i.e.  of $\sql$. This 
effectively  inverts the sign of
$\c_2$.}

According to  standard  discussions
of the appearance  of similar asymptotic series in QM and QFT  problems 
 this seems to  suggest  that 
the string background we are expanding  about is
  actually  unstable, despite its apparent stability under 
  small  fluctuations of string coordinates (all fluctuation modes in section 3 had 
  non-negative values of  squares of their masses).\foot{The standard argument \cite{dyson} 
  based on $e^2 \to - e^2$  continuation that makes perturbative
  vacuum unstable implies  that the complex 
coupling constant space exhibits a cut on the negative real axis
and consequently the expansion  in small $e$ 
near a perturbatively stable vacuum  should lead to an 
 asymptotic sign-alternating  and thus Borel summable series. 
 In the case when  the coefficients in the series are not sign-alternating, 
  the Borel transformed series no longer converges, i.e. the perturbation series 
 is  not Borel-summable. From the standpoint of 
the argument of \cite{dyson} this case appears to 
 correspond to developing 
 perturbation theory around an unstable vacuum state.} 
The instability should manifest itself in the  existence of 
complex energies  but  it is 
 not  obvious what might be the origin of this  instability,

 Absence of Borel summability of 
 perturbative expansions
occurs in
all quantum-mechanical potential problems
in which the expansion is done near  a local (but not global) minimum 
of the potential \cite{vainshtein, benderwu}.\foot{In  QFT context, 
 ref.\cite{lipatov} 
 argued that the large order
 behavior of coefficients  of
perturbation theory around a stable vacuum state may be explained  by the  
existence  of a  classical euclidean solution of the effective 
action appearing at each order of perturbative expansion
(a complex instanton).
In \cite{BGZ} it was argued that the lack of 
 Borel summability of perturbation theory near an unstable vacuum 
 may 
be related to 
 the existence of a ``real instanton'' of the original action.}
By analogy with such systems we may 
interpret this  apparent  non-summability of the string $\a' \sim { 1 \ov \sql}$ 
perturbation 
theory as a signal that  the closed spinning string (or, equivalently, 
 the null cusp solution) 
is  only  a local minimum  of the \adss superstring  action. 
If this is indeed the correct interpretation, there should be a 
tunneling solution  connecting  it  to some  ``global vacuum'' 
state.

 
  It is interesting to note that 
the details of our 1-loop and 2-loop calculations (cf. \rf{onn},\rf{hk})
 suggest that the perturbation
 theory of the {\it bosonic} \adss sigma model leads  
 to a
  sign-alternating -- and hence 
 Borel-summable -- series 
  while the addition of the fermionic contributions 
 spoils  this feature.\foot{This might be an  indication 
that  the ``global vacuum''  may  involve some nontrivial classical 
profiles for the fermionic  fields.}

A non-resummable perturbation theory expansion is ambiguous 
in the sense that either the function 
being expanded has indeed 
a cut along the real axis in the coupling plane 
 (in which case the expansion is
 meaningless) or the singularity is cancelled by terms whose derivatives vanish 
at the expansion point \cite{BGZ}.
What happens in the present case, both from the 
Bethe ansatz  and the string theory 
points of view,  remains to be clarified.

\bigskip 

One generalization of the 2-loop superstring  computation described in this paper 
is to the case  of null cusp solution with non-zero angular momentum $J$ in $S^5$. 
This   in particular may  allow one  to verify  the  
 suggestion  \ci{am2} that certain
 terms in the corresponding  anomalous dimension found in the limit 
  when ${J \ov \sql \ln S } \ll 1$ 
 are determined only by the bosonic $S^5$  contributions.
The relevant solution with non-zero $J$ is (cf. \rf{bc},\rf{ch}, \ci{krtt}; see also  footnote
15 above)
\be
&& {\bar z}= \sqrt 2   e^{-\kappa \,\uo + \ut}\ , \ \ \ \ \ \  \ \ \ \   \varphi=\nu'\,\uo \ , \ \ \ \ \ 
  \ \ \ \ \  \kappa  = \sqrt{ 1 - \nu'^2} \ , 
  \cr
&&{\bar x}^0=e^{- \kappa \,\uo + \ut }\cosh(   \ut + \kappa \uo)\ , \ \ \ \ \ \ \ \ \ \ 
{\bar x}^1=  e^{- \kappa \,\uo + \ut }\sinh( \ut + \kappa \uo)\ . \la{hok}
\ee
%
%
Here $\varphi$ is an angle of $S^5$  and $\nu'= i \nu$, where $ \ J= \sql \nu$
is the angular momentum of the corresponding  spinning string background 
with Minkowski world sheet.   
A technical complication is 
that, unlike the case of $\nu'=0$ discussed above, 
the denominator of the bosonic propagator  no longer has a  Lorentz-covariant form. 
Consequently, the direct calculation of momentum integrals 
becomes quite cumbersome. Moreover, the existence 
of fluctuation fields of mass proportional to $\nu'$ requires that their
contribution is  treated exactly. We  leave this problem for the future.

\section*{Acknowledgments }

We are  grateful to  A. Belitsky, 
G. Korchemsky, M. Kruczenski, R. Metsaev,  A. Pivovarov, M. Staudacher     and  A. Vainshtein 
for useful communications and discussions. 
We thank G. Korchemsky for sending us a copy of 
\ci{bkk} prior to its publication. 
R.R.   acknowledges
the support of the National Science Foundation under grant
PHY-0608114  and of a Sloan Research Fellowship. 
 A.A.T. acknowledges  the support of
the PPARC, INTAS 03-51-6346, EC MRTN-CT-2004-005104   and the RS
Wolfson award. Part of this work was done
while R.R. was a participant of the 
"Advancing Collider Physics: from Twistors to Monte Carlos" 
workshop at the Galileo Galilei Institute for Theoretical Physics in Florence.
\bigskip

\renewcommand{\theequation}{A.\arabic{equation}}
\renewcommand{\thesection}{A}
 \setcounter{equation}{0}
\setcounter{section}{1} \setcounter{subsection}{0}

 \section*{Appendix:  Details of  fluctuation Lagrangian 
 }
The relation between the Poincare coordinate fields $z, x^m$ and the fluctuation fields 
in \rf{bf} is as follows ($\eps$ is a formal  expansion parameter that should be set to 1
at the end)
\be
z&=&\frac{e^{-\sqrt 2 \ n_1\cdot\uu -\eps \tfpt}}
{\sin(\frac{\pi}{4}+\eps\tphi))\sqrt{1+\eps^2(\xi^2+\eta^2)}} \ , 
\cr
x^0&=&e^{-\sqrt 2\ n_1\cdot\uu -\eps \tfpt}
\cosh(\sq \ n_2\cdot \uu -\eps\tfpo)
\cot(\frac{\pi}{4}+\eps\tphi) \ , 
\cr
x^1&=&-e^{-\sq \ n_1\cdot\uu -\eps \tfpt}
\sinh(\sq \ n_2\cdot \uu -\eps\tfpo)
\cot(\frac{\pi}{4}+\eps\tphi)
\cr
x^2&=&\frac{\eps\ e^{-\sq\ n_1\cdot\uu -\eps \tfpt}\ \eta}
{\sin(\frac{\pi}{4}+\eps\tphi)\sqrt{1+\eps^2(\xi^2+\eta^2)}}\ , 
\cr
x^3&=&\frac{\eps\ e^{-\sq\ n_1\cdot\uu -\eps \tfpt}\ \xi}
{\sin(\frac{\pi}{4}+\eps\tphi)\sqrt{1+\eps^2(\xi^2+\eta^2)}} \ . 
\ee
Theexpansion of the  matrix \rf{baa},\rf{nu} 
that enters the action \rf{fii}  has the form  
\be
{\cal N}_{ab}=\eta_{ab} +\eps
\begin{pmatrix}
2\tphi+\tfpt
-\frac{1}{\sqrt{2}}(n_1 \cdot \partial\tphi +n_1 \cdot \partial\tfpt)&
\tfpo-\frac{1}{\sqrt{2}}n_1 \cdot \partial \tfpo \cr
-\tfpo-\sqrt{2} n_2 \cdot \partial\tphi-\frac{1}{\sqrt{2}}n_2 \cdot \partial \tfpt
&
-2\tphi-\tfpt -\frac{1}{\sqrt{2}}n_2 \cdot \partial\tfpo
\end{pmatrix}_{ab}
+{\cal O}(\eps^2) \  . 
\ee
The  bosonic propagator corresponding to the quadratic part of the 
   Lagrangian \rf{bf}  is 
\be \la{bopr}
K^{-1}_B(q)=
\begin{pmatrix}
  \frac{-1}{2(q^2+4)}
& \frac{i \sq\ n_2\cdot q}{q^2\,(q^2+4)}
& \frac{i \sq\ n_1\cdot q}{q^2\,(q^2+4)}
& 0 & 0  & 0 \cr
  \frac{-i\sq\  n_2\cdot q}{q^2\,(q^2+4)}
& \frac{(q^2)^2+ 4  (n_1\cdot q)^2}{(q^2)^2(q^2+4)}
& \frac{-4  n_1\cdot q\,n_2\cdot q}{(q^2)^2(q^2+4)}
& 0 & 0  & 0 \cr
  \frac{-i \sq\ n_1\cdot q}{q^2\,(q^2+4)}
& \frac{-4  n_1\cdot q\,n_2\cdot q}{(q^2)^2(q^2+4)}
& \frac{(q^2)^2+ 4 (n_2\cdot q)^2}{(q^2)^2(q^2+4)}
& 0 & 0  & 0 \cr
  0 & 0  & 0
& \frac{1}{2(q^2+  2 )}
& 0 & 0  \cr
  0 & 0  & 0 & 0
& \frac{1}{2(q^2+2)} \cr
  0 & 0  & 0 & 0 & 0
& \frac{1}{2\,q^2}\,\id_5
\end{pmatrix}
\ee
The fermionic propagator corresponding to \rf{pr} is  
%
%
%
\be\label{fpr}
K_F^{-1}=\frac{\Gamma_L}{q^2+1 }\Big[
 i n_1\times q(\Gamma^0-\sqrt{2}\Gamma^9)
- i n_2\times q\Gamma^1
+
 n_1\times n_2\Gamma^{019}
\Big]{\cal C}^{-1} \ , 
\ee
where $n \times  q = \eps^{ab} n_a q_b$,  
$\Gamma_L=\ha (  1 + \G_{11})$ is the left-handed chiral projector
and  ${\cal C}= \G^0$ is the charge conjugation matrix (for notation see also
\ci{rtt}).




\end{document}